\documentclass[a4paper, 11pt]{article}   
\usepackage[T1]{fontenc}
\usepackage[utf8]{inputenc}		
\usepackage[english]{babel} 
\usepackage{geometry}
\newgeometry{a4paper, top=35mm, left=27mm, right=27mm, bottom=33mm, headsep=10mm, footskip=12mm} 

\usepackage{sectsty}
\paragraphfont{\footnotesize}

\usepackage{siunitx} 
\usepackage{amsmath}
\usepackage{dsfont} 
\usepackage{booktabs}
\usepackage[normal, font=footnotesize]{caption} 
\usepackage{subcaption}

\usepackage[onehalfspacing]{setspace} 
\usepackage{textgreek} 
\usepackage{float} 
\usepackage[nolist]{acronym} 

\usepackage{amssymb} 
\usepackage{pgfplots}
\usepackage{epstopdf}
\usepackage{graphicx} 
\graphicspath{{./Figures/}}
\usepackage{color} 

\usepackage{lmodern} 

\usepackage[colorlinks]{hyperref}  

\definecolor{darkblue}{rgb}{0.0, 0.15, 0.45}
\hypersetup{colorlinks,linkcolor=black,citecolor=darkblue,urlcolor=black}

\usepackage[super, comma, sort&compress]{natbib}  

\usepackage{icomma} 

\usepackage{titlesec}
\definecolor{gray75}{gray}{0.75} 

\usepackage{fancyhdr} 
\fancypagestyle{plain}{
\fancyhf{}
\fancyhead[LE]{\nouppercase{\leftmark}} 
\fancyhead[RO]{\nouppercase{\rightmark}} 
\fancyfoot[LE, RO]{\thepage} 
\fancyheadoffset{0cm} 
}
\fancyheadoffset{0cm} 

\usepackage[bottom,hang]{footmisc} 
\usepackage{placeins}

\usepackage{upgreek}

\begin{document}

\begin{acronym}
	\acro{GFET}{graphene field effect transistor}
	\acro{FET}{field effect transistor}
	\acro{cQD}{colloidal quantum dot}
	\acro{AFM}{atomic force microscope}
	\acro{L}{channel length}
	\acro{W}{channel width}
	\acro{EHD}{electrohydrodynamic nanoprinting}
	\acro{IR}{infrared}
	\acro{SWIR}{short wave infrared}
	\acro{PSD}{power spectral density}
	\acro{EDT}{ethane-1,2-dithiol}
	\acro{LbL}{layer-by-layer}
	\acro{NEI}{noise equivalent irradiance}
	\acro{NEP}{noise equivalent power}
	\acro{LSQ}{least squares fit}
	\acro{TLM}{transfer length method}
\end{acronym}

\pagenumbering{arabic}
\setcounter{page}{1}

\section*{Scaling of Hybrid QDs-Graphene Photodetectors to Subwavelength Dimension}
\label{chapter_scaling}


Gökhan Kara\textsuperscript{a}, Patrik Rohner\textsuperscript{a,b}, Erfu Wu\textsuperscript{a}, Dmitry N. Dirin\textsuperscript{b,d}, Roman Furrer\textsuperscript{a}, Dimos Poulikakos\textsuperscript{c}, Maksym V. Kovalenko\textsuperscript{b,d}, Michel Calame\textsuperscript{a,e}, Ivan Shorubalko\textsuperscript{a,*}\\
~\\
\begin{scriptsize}
	(a) Laboratory for Transport at Nanoscale Interfaces, Empa – Swiss Federal Laboratories for Materials Science and Technology, CH-8600 Dübendorf, Switzerland~\\
	(b) Department of Chemistry and Applied Biosciences, ETH – Swiss Federal Institute of Technology Zurich, CH-8093 Zurich, Switzerland~\\
	(c) Laboratory of Thermodynamics in Emerging Technolohgies, Department of Mechanical and Process Engineering, ETH – Swiss Federal Institute of Technology Zurich, CH-8092 Zurich, Switzerland~\\
	(d) Laboratory for Thin Films and Photovoltaics, Empa – Swiss Federal Laboratories for Materials Science and Technology, CH-8600 Dübendorf, Switzerland~\\
	(e) Department of Physics and Swiss Nanoscience Institute, University of Basel , CH-4056 Basel, Switzerland~\\
	* Email: ivan.shorubalko@empa.ch
\end{scriptsize}
~\\

\begin{footnotesize}
	\paragraph{Abstract} ~\\
	Emerging \acf{cQD} photodetectors currently challenge established state-of-the-art \acl{IR} photodetectors in response speed, spectral tunability, simplicity of solution processable fabrication, and integration onto curved or flexible substrates. Hybrid phototransistors based on 2D materials and \acsp{cQD}, in particular, are promising due to their inherent photogain enabling direct photosignal enhancement. The photogain is sensitive to both, measurement conditions and photodetector geometry. This makes the cross-comparison of devices reported in the literature rather involved. Here, the effect of device length $L$ and width $W$ scaling to subwavelength dimensions (sizes down to \qty{500}{nm}) on the photoresponse of graphene-PbS \acs{cQD} phototransistors was experimentally investigated. Photogain and responsivity were found to scale with $1/LW$, whereas the photocurrent and specific detectivity were independent of geometrical parameters. The measurements were performed at scaled bias voltage conditions for comparable currents. Contact effects were found to limit the photoresponse for devices with $L<3$ \unit{\um}. The relation of gate voltage, bias current, light intensity, and frequency on the photoresponse was investigated in detail, and a photogating efficiency to assess the \acs{cQD}-graphene interface is presented. In particular, the specific detectivity values in the range between $10^8$ to $10^9$ \unit{Jones} (wavelength of \qty{1550}{nm}, frequency \qty{6}{Hz}, room temperature) were found to be limited by the charge transfer across the photoactive interface.
\end{footnotesize}
~\\

\begin{footnotesize}
	\noindent \textbf{Keywords:} colloidal quantum dots, graphene, phototransistors, infrared photodetectors, scaling, subwavelength
\end{footnotesize}

\subsection*{Introduction}

Emerging photodetectors based on \acfp{cQD} are currently challenging state-of-the-art \acf{IR} photodectors by low cost solution processing\cite{Kovalenko2015,Lu2019}, simple spectral tunability\cite{Kovalenko2015,Lu2019}, multiband absorption\cite{Tang2019a,Tang2019b,Tang2020}, response speed\cite{Gao2016a,Biondi2021,Vafaie2021}, and integration onto flexible substrates\cite{Bessonov2017,Tang2019,Polat2019,Kim2023a}. PbS \acsp{cQD} \acs{SWIR} photodiode arrays were recently fabricated with a pixel pitch of \qty{1.62}{\um}.\cite{JonathanSteckel2021} This is about five times smaller than the pixel size of common InGaAs photodetectors. Reduced dimension are, on the contrary, also limiting the number of photons reaching a detector. The integration of photonic structures is, thus, an attractive way to enhance light-matter interaction in \acs{cQD} photodetectors.\cite{Greboval2021a,Dang2022} Another way to overcome low light levels is by coupling a light absorber with a transistor. By introducing PbS \acsp{cQD} as a photogate to a \acf{GFET}, photoresponsivities of $10^6$ to $10^7$ \unit{A \per W} were demonstrated.\cite{Konstantatos2012,Sun2012} The high responsivity values were also reflected in the specific detectivity of $10^{13}$ \unit{Jones} at room temperature\cite{Konstantatos2012} that is comparable to commercial InGaAs photodetectors.

Research in hybrid phototransistors is focused on reducing dark currents by exchanging the transistor channel material with TMDCs\cite{Kufer2015,Kufer2016b,Ozdemir2019,Kundu2021} or metal oxides\cite{Choi2020,Kim2023a}, energy barrier engineering\cite{Kufer2016b,Bessonov2017,Ahn2020,Ahn2022}, longer charge extraction distance in the \acs{cQD} films\cite{Nikitskiy2016}, extending the spectral range to longer wavelength \cite{Grotevent2021a,Ni2017,Kundu2021}, or technological applicability to flexible substrates\cite{Bessonov2017,Polat2019,Kim2023a} and printing\cite{Grotevent2019,Grotevent2021a,Kara2023}. Although it is commonly reported that the photogain $G_{ph}$ follows a $1/L^2$ dependence in those devices\cite{Koppens2014,Saran2016,Huo2018}, limited effort has been spent to experimentally validate this prediction. In addition, the photoresponse is highly dependent on measurement conditions (e.g., bias, illumination), geometry, and batch-specific fabrication. Thus, reproducibility and the cross-comparison of individual reports remain highly involved, and derived scaling laws are inconclusive.

Here, we experimentally demonstrate the scaling of hybrid graphene - PbS cQD phototransistors with channel length, width, and \acs{cQD} film thickness. We highlight a $1/LW$ dependence of the photogain and responsivity, and a dimension-independent photocurrent and specific detectivity at scaled bias voltage condition for comparable currents. We further show the photoresponse's gate voltage, bias current, light intensity, and frequency dependency and derive a photogating efficiency for a \acs{cQD}-graphene interface evaluation.

\subsection*{Results and Discussion}

\subsubsection*{Photocurrent Relation of Hybrid Phototransistors}

\begin{figure}[h!tb] 
	\centering
	\includegraphics[width=\linewidth]{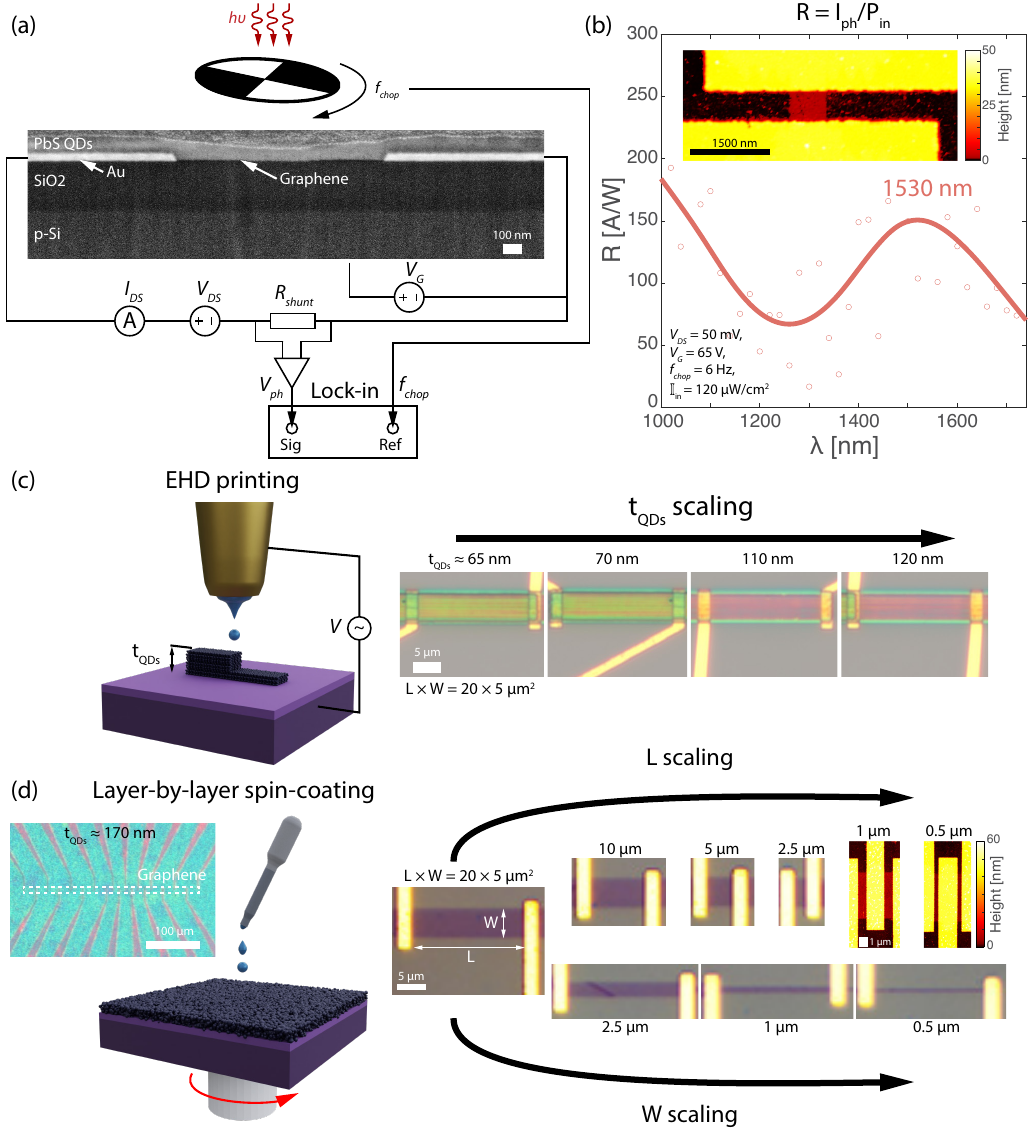}
	\caption{Photodetector characterization scheme and used PbS \acs{cQD} deposition techniques. (a) the cross-section of a hybrid graphene-PbS \acs{cQD} phototransistor. Two Au electrodes contacted the graphene channel, and \qty{285}{nm} SiO\textsubscript{2} separated the p-Si back gate from the channel. A $V_{DS}$ bias was applied, and $V_G$ on the back gate modulated the 2D charge carrier density $n$ in the channel. The incoming light with energy $E_{ph}=hc/\lambda$ was chopped at a frequency $f_{chop}$. Photocurrents were extracted with a lock-in amplifier measuring $V_{ph}$ over a shunt resistance. (b) Photoresponsivity $R$ of a subwavelength dimensioned photodetector of $0.5\times0.5$ \unit{\um^2}. The inset shows the AFM image of the \acs{GFET} before \acs{cQD} deposition. (c) \acs{EHD} printing illustration and photodetectors with a channel geometry $L\times W$ of $20\times5$ \unit{\um^2}. The PbS \acs{cQD} thickness was varied. (d) \acs{LbL} spin coating illustration. The inset on the left shows an optical microscope image after PbS \acs{cQD} fabrication. The L- and W-scaling devices are shown before \acs{cQD} deposition (optical microscope and AFM).}
	\label{fig:ch_Scaling_Figure1}
\end{figure}

Figure \ref{fig:ch_Scaling_Figure1} (a) shows a cross-section of a hybrid phototransistor. Two gold electrodes (source and drain) contact the graphene channel, and a p-Si back gate enables charge carrier density tuning in the channel by the field effect. The phototransistor has a channel length (L), width (W), and a \acsp{cQD} film thickness $t_{QDs}$ as the photogate on top. The incoming light creates electron-hole (e-h) pairs in the absorbing \acs{cQD} layer. The Fermi level alignment between graphene and the \acs{cQD} film causes an electric field at the interface (depletion region) that separates the photo-generated e-h pairs and leads to the commonly observed hole transfer to the graphene channel. The electrons stay trapped in the film for a time $\tau_{trap}$, gating the channel.\cite{Konstantatos2012} The photoinitiated e-h separation current between the \acsp{cQD} film and graphene is small compared to the channel current $I_{DS}$ and is thus neglected. The photogate modulation of $I_{DS}$ is an enhancement of the photosignal and is described as photogain by the rate equation
\begin{equation}
	G_{ph} = \frac{\# \; charges \; observed}{\# \; photogenerated \; charges} =\frac{1/\tau_{transit}}{1/\tau_{trap}}.
\end{equation}
$\tau_{transit}$ is the transit time of a charge carrier in the channel between source and drain contacts.\cite{Koppens2014,Fang2017,Huo2018} Within the Drude model, the transit time can be stated by bias condition and geometry of the channel and leads to
\begin{equation}
	\label{eq:ch_Scaling_GainVds}
	G_{ph} = \frac{1}{L^2}\tau_{trap}\mu V_{DS}.
\end{equation}
$V_{DS}$ is the bias voltage between source and drain, and $\mu$ the charge carrier mobility in the channel. The photocurrent of the hybrid phototransistors can thus be described by
\begin{equation}
	\label{eq:ch_Scaling_IphGen}
	I_{ph} = G_{ph}\phi\eta_{Gph} e,
\end{equation}
where $e$ the elementary charge, and $\eta_{Gph}$ is a photogating efficiency describing how likely an incoming photon is creating a charge photogating the transistor channel.\cite{Saleh2007,Koppens2014} The photonflux $\phi = P_{in}/E_{ph}$ is defined by the incoming light power $P_{in}$ and photon energy $E_{ph}=hc/\lambda$. $h$ is Planck's constant, $c$ the speed of light, and $\lambda$ the wavelength of the incoming light. Thus, the photoresponsivity
\begin{equation}
	\label{eq:ch_Scaling_R}
	R = \frac{I_{ph}}{P_{in}},
\end{equation}
is proportional to the $G_{ph}$ and often stated as $R\sim1/L^2$.

Figure \ref{fig:ch_Scaling_Figure1} (a) also shows the experimental setup used to characterize the photoresponse of the investigated detectors. A bias voltage $V_{DS}$ was applied between the source and drain contacts, and the current $I_{DS}$ was measured simultaneously. A gate voltage $V_{G}$ was applied to the back gate to control the 2D charge carrier density $n = C_{SiO_2}V_G/e$ in the graphene channel, where $C_{SiO_2}\approx 11.9$ \unit{nF\per cm^2} is the capacitance of the gate oxide. The incoming light was chopped at a frequency $f_{chop}$. This light modulation induced an AC photocurrent on top of $I_{DS}$ that was extracted as a voltage drop $V_{ph}$ over a shunt resistance. The incoming light power was characterized by a reference detector. The calculated the photoresponsivity for a sub-wavelength dimensioned device of $0.5 \times 0.5$ \unit{\um^2} is shown in Figure \ref{fig:ch_Scaling_Figure1} (b). The inset shows the AFM image of the characterized device before \acs{cQD} sensitization. A photoresponsivity of \qty{150}{A\per W} was reached at the first excitonic peak at \qty{1530}{nm}. The spin-coating of the PbS \acsp{cQD} lead to a blue shift of about \qty{70}{nm} compared to their excitonic peak at \qty{1600}{nm} (diameter $\sim$ \qty{6}{nm})\cite{Moreels2009} after synthesis. All measurements were performed in vacuum and at room temperature.

Further, \acf{EHD} and \acf{LbL} spin coating were compared to assess the influence of \acs{cQD} deposition techniques on the photoresponse. First, in Figure \ref{fig:ch_Scaling_Figure1} (c), the deposition of \acsp{cQD} by EHD printing is depicted. By applying an AC voltage between the gold coated nozzle and the substrate, \acsp{cQD} accumulate at the formed meniscus, and droplets markedly smaller than the nozzle diameter are pulled out by the electric field from the apex of the meniscus. This enables printing features with a resolution in the \qty{100}{nm} range.\cite{Galliker2012,Onses2015} For PbS \acsp{cQD} in particular, a resolution of $\sim$ \qty{1}{\um} was previously demonstrated.\cite{Grotevent2019} This method enables precise spatial control of placing \acsp{cQD} in a desired geometrical pattern. After printing, a solid-state ligand exchange treatment was performed to substitute the native oleic acid with \acl{EDT} ligands. This single-step treatment reduces the interparticle distance, a $t_{QDs}$ shrinkage of about 40\% was observed, leading to conductive films. Additionally, the extra control of film thickness by \acs{EHD} printing on a single sample level allowed to vary $t_{QDs}$ between 65 to \qty{120}{nm}. The geometry of the device channels were $L\times W = 20 \times 5$ \unit{\um^2}.

The second applied method was the \acf{LbL} spin coating. A thin layer of \acsp{cQD} was spin coated and directly treated with a solid-state ligand exchange procedure to \acs{EDT}. These two steps were repeated six times, resulting in $t_{QDs}$ of \qty{170}{nm}. By this method, devices with different $L$ ranging from 20 down to 0.5 \unit{\um} and a $W=5$ \unit{\um} were sensitized with \acsp{cQD}. In addition, the $W$ scaling was investigated by changing the width between 5 down to 0.5 \unit{\um} with a fixed $L=20$ \unit{\um}. Figure \ref{fig:ch_Scaling_Figure1} (d) shows the investigated devices with the \acs{LbL} spin-coating approach before deposition. In contrast to printing, the entire sample was covered with one \acs{cQD} thickness (inset). The \acs{LbL} spin coating has been proven to yield dense and crack-free conductive films\cite{Luther2008a}, optimally suited for solar cells and photodetectors.\cite{Chuang2014,Albaladejo-Siguan2021a,Wu2023}

\begin{figure}[h!tb] 
	\centering
	\includegraphics[width=\linewidth]{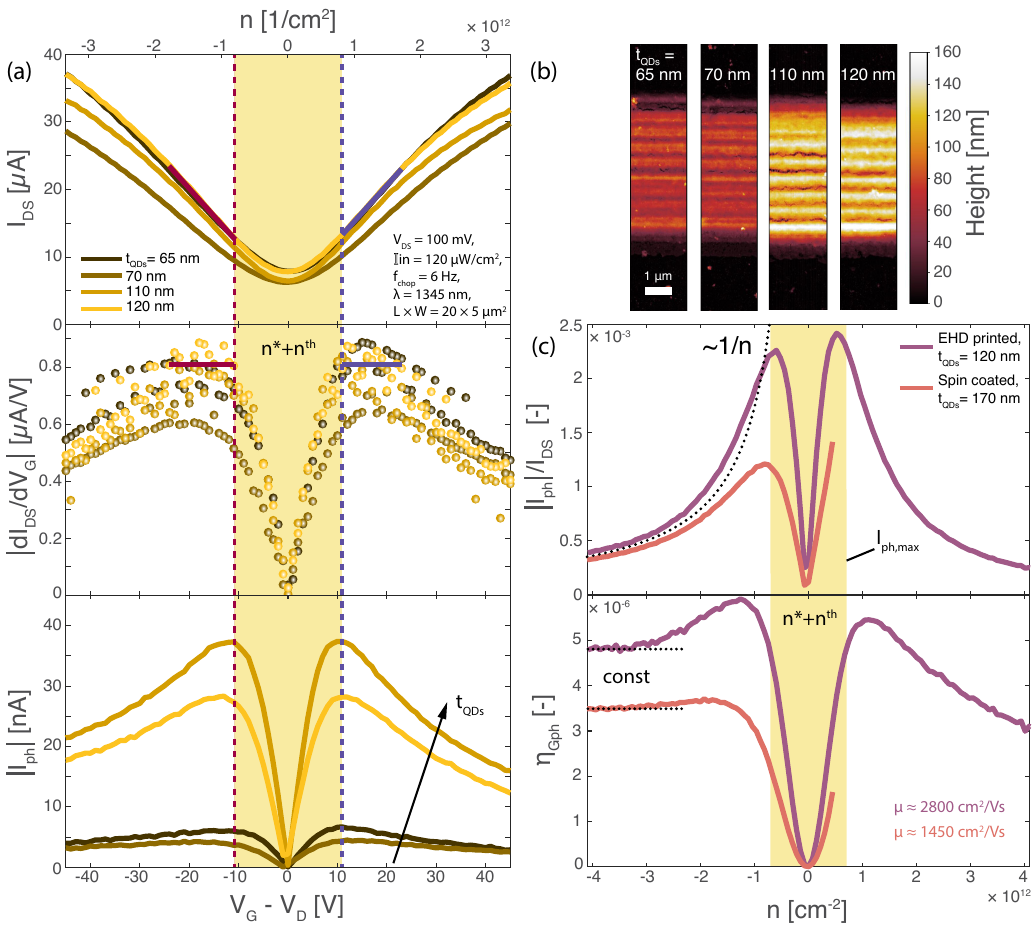}
	\caption{Photoresponse transfer characteristic and comparison of devices with \acs{EHD} and \acs{LbL} spin-coated films. (a) transfer characteristic of four devices with \acs{EHD} PbS \acs{cQD} film of different thicknesses. From top to down, the source-drain current $I_{DS}$, transconductance $g_m = dI_{DS}/dV_G$ and photocurrent $I_{ph}$ from the same measurement were extracted. The gate voltage $V_G$ is corrected by the Dirac point voltage $V_D$. $g_m$ was derived from $I_{DS}$ by a numerical derivative and is highlighted as dots to distinguish from the measured quantities. The dashed lines show the maximum $I_{ph}$ in the p- (red) and n-doped (blue) regions of graphene. The solid red and blue line in $g_m$ and $I_{DS}$ indicates the region of about constant maxima of $g_m$ and is displayed for the device with $t_{QDs} = 120$ \unit{nm} only. (b) AFM images of the characterized devices in (a), showing the thickness $t_{QDs}$ of \acs{cQD} films on top of graphene. (c) cross-comparison of two devices with \acs{EHD} printed and \acp{LbL} spin coated \acp{cQD} films. $I_{ph}$ (top panel) was normalized by $I_{DS}$ to account for contact resistance and bias conditions $V_{DS}$ of 100 (\acs{EHD}) and 200 \unit{mV} (\acs{LbL}). Devices were measured at a wavelength $\lambda$ matching the respective first excitonic peaks of 1345 (\acs{EHD}) and 1550 \unit{nm} (\acs{LbL}), and $\mathds{I}_{in} = 120$ \unit{\mu W \per cm^2}. The bottom panel shows the two devices' derived photogating efficiency $\eta_{Gph}$. Yellow shading illustrates where the highest photocurrent $I_{ph}$ has been found. The shaded region might originate from random potential in graphene around the Dirac point $n^*+n^{th}$, where a gate can not control the conductivity of graphene.}
	\label{fig:ch_Scaling_Figure2}
\end{figure}

Figure \ref{fig:ch_Scaling_Figure2} (a) shows the characterized devices of EHD-printed \acsp{cQD} films with different layer thicknesses. From top to down, the plots show the source-drain currents $I_{DS}$, transconductance $g_m = dI_{DS}/dV_G$ and the amplitude of the photocurrent $\big| I_{ph} \big|$ while the gate voltage $V_G$ was swept. $g_m$ was derived by a numerical derivative from $I_{DS}$. The measurements were performed at a bias voltage $V_{DS}$ of \qty{100}{mV}, irradiance $\mathds{I}_{in}$ of \qty{120}{\micro W \per cm^2}, chopping frequency $f_{chop}$ of \qty{6}{Hz}, and the curves were corrected by the individual Dirac point voltage $V_D$ (charge neutrality point) of the \acsp{GFET} (average $V_D$ of \qty{-20}{V}). The first excitonic peak of the \acsp{cQD} shifted to a wavelength $\lambda$ of \qty{1345}{nm} (shift of about 250 nm). The blue shift of the absorption peak might be explained by oxidation of the \acsp{cQD} during printing. The maximum of the measured photocurrent $I_{ph}$ was found at the left, hole-doped region $p$, and right, electron-doped region $n$, of the Dirac point. The dashed lines (red and blue) indicate where the maximum photocurrents were measured that, on average, corresponds to 10 V ($n\approx7\times10^{11}$ \unit{1/cm^2}) from the Dirac point.

The solid red and blue lines (p- and n-doped region) in the $g_m$ and $I_{DS}$ plots indicate where $g_m$ reaches about a constant maximum, as illustrated for the device with $t_{QDs} = 120$ \unit{nm}. The maxima of the photocurrent coincide with this region and thus are found where the transconductance is the highest, as reported previously.\cite{Huang2013} The yellow shaded region of Figure \ref{fig:ch_Scaling_Figure2} (a) indicates the influenced region by random potential fluctuations around the Dirac point in graphene. In this region, the applied gate voltage loses its ability to tune $n$ up to a point where charges are only redistributed locally (e-h charge puddles) and, thus, does not affect the conductivity anymore. This causes a drop of $g_m$ towards the Dirac point and depends on residual doping and temperature $n^*+n^{th}$.\cite{Du2008}

Figure \ref{fig:ch_Scaling_Figure2} (b) depicts the four devices' AFM images after ligand exchange treatment. Thicker \acs{cQD} films absorb more light reaching 100\% absorption at about \qty{500}{nm} for an absorption coefficient of $\sim 1\times10^{5}$ \unit{\per cm}.\cite{Moreels2009} Correspondingly, an increasing trend of the photocurrent with thicker $t_{QDs}$ films is observed in the bottom panel of Figure \ref{fig:ch_Scaling_Figure2} (a). However, to extract charges from the \acs{cQD} layer, they need to be in the depletion region close to the interface.\cite{Nikitskiy2016} This is why a saturation of the photocurrent is expected before $t_{QDs}\approx500$ \unit{nm} is reached. In a previous study, this saturation thickness was found to be about \qty{160}{nm} for EHD-printed \acs{cQD} films.\cite{Grotevent2019}

The photoresponsivity $R$ is used to assess detectors' efficiency in converting light into a photocurrent. However, $R$ is a derived quantity and describes the proportionality between the photocurrent and the incoming light power.\cite{Saleh2007} Thus, in this study, the main focus was put on $I_{ph}$ as the measured quantity at compatable conditions across the investigated devices. Also, applying a constant bias voltage might not always be applicable over the different length scales as the current is proportionally increasing with $1/L$. A comparison should thus aim to keep the applied electric field $\big|\vec{E}\big| = V_{DS}/L$ constant, which induces a constant $I_{DS}$ condition. Therefore, equation (\ref{eq:ch_Scaling_GainVds}) can be reexpressed with a bias-current, and further taking the channel conductivity ($\sigma=en\mu$) into account, yields
\begin{equation}
	\label{eq:ch_Scaling_GainIds}
	G_{ph} = \frac{1}{LW} \frac{1}{en} \tau_{trap} I_{DS}.
\end{equation}
Likewise, devices observe the same irradiance $\mathds{I}_{in}$ rather than the same light power $P_{in}$. Including (\ref{eq:ch_Scaling_GainIds}) and $\mathds{I}_{in}$ into equation (\ref{eq:ch_Scaling_IphGen}) hence results in
\begin{equation}
	\label{eq:ch_Scaling_Iph}
	I_{ph} = \frac{1}{E_{ph}}\frac{1}{n}\eta_{QE}\tau_{trap} \mathds{I}_{in} I_{DS},
\end{equation}
and predicts a geometry-independent photocurrent. Note that $n$ states the charge carrier density for electrons if positive and for holes if negative to account for the ambipolar graphene FET behavior.

In Figure \ref{fig:ch_Scaling_Figure2} (c), the photocurrent of two devices close to the photocurrent saturation limited film thickness were compared. The two devices were fabricated independently, and the \acsp{cQD} deposition method was varied between \acs{EHD} printing and \acs{LbL} spin coating as described in Figure \ref{fig:ch_Scaling_Figure1} (b) and (c). $I_{ph}$ was normalized by $I_{DS}$ to account for contact resistances. The photoresponse shows a $1/n$ proportionality, and drops to zero at the Dirac point of graphene.

Although solid-state ligand exchanged \acs{cQD} films are predominantly fabricated by a \acs{LbL} approach in the literature, they do not necessarily result in a higher conductivity than a single-step ligand exchanged approach.\cite{Luther2008a} Accordingly, the devices with the EHD printed \acs{cQD} films (single-step ligand exchange) demonstrate a slightly higher photoresponse of about a factor of 1.8. The ratio of the mobilities $\mu$ is about 1.9. Using equation (\ref{eq:ch_Scaling_Iph}), the photogating efficiency $\eta_{Gph}$ was estimated from the experiments as shown in the bottom panel. The extracted $\eta_{Gph}$ also takes the different wavelengths from the measurements (different first excitonic peaks, although the same \acs{cQD} are used for the different devices) into account. The efficiency ratio is about 1.6 and similar to the $I_{ph}$ and $\mu$ ratios. The photogating efficiency can further be expressed as
\begin{equation}
	\eta_{Gph} = \eta_{trans} \eta_{abs},
\end{equation}
where $\eta_{trans}$ describes the charge transfer efficiency, how well a photogenerated charge can be extracted from the \acs{cQD} layer to graphene, and $\eta_{abs}$ light absorption efficiency, how many e-h pairs are generated per incident photon.\cite{Koppens2014,Fang2017,Dorodnyy2018} Furthermore, $\eta_{abs}$ can be estimated to be in the range of 0.6 to 0.8. As the magnitude of $\eta_{Gph}$ is in the order of $10^{-6}$ and one order of magnitude is lost to the absorption ($\eta_{abs} < 1$), there is only about one every $10^5$ charge carriers that are extracted to graphene. Thus $\eta_{trans}$ is a major source for improvement and stresses that a single-step ligand exchange is not a limiting factor.

\subsubsection*{Scaling of Pixel Dimensions}

\begin{figure}[h!tb] 
	\centering
	\includegraphics[scale=0.725]{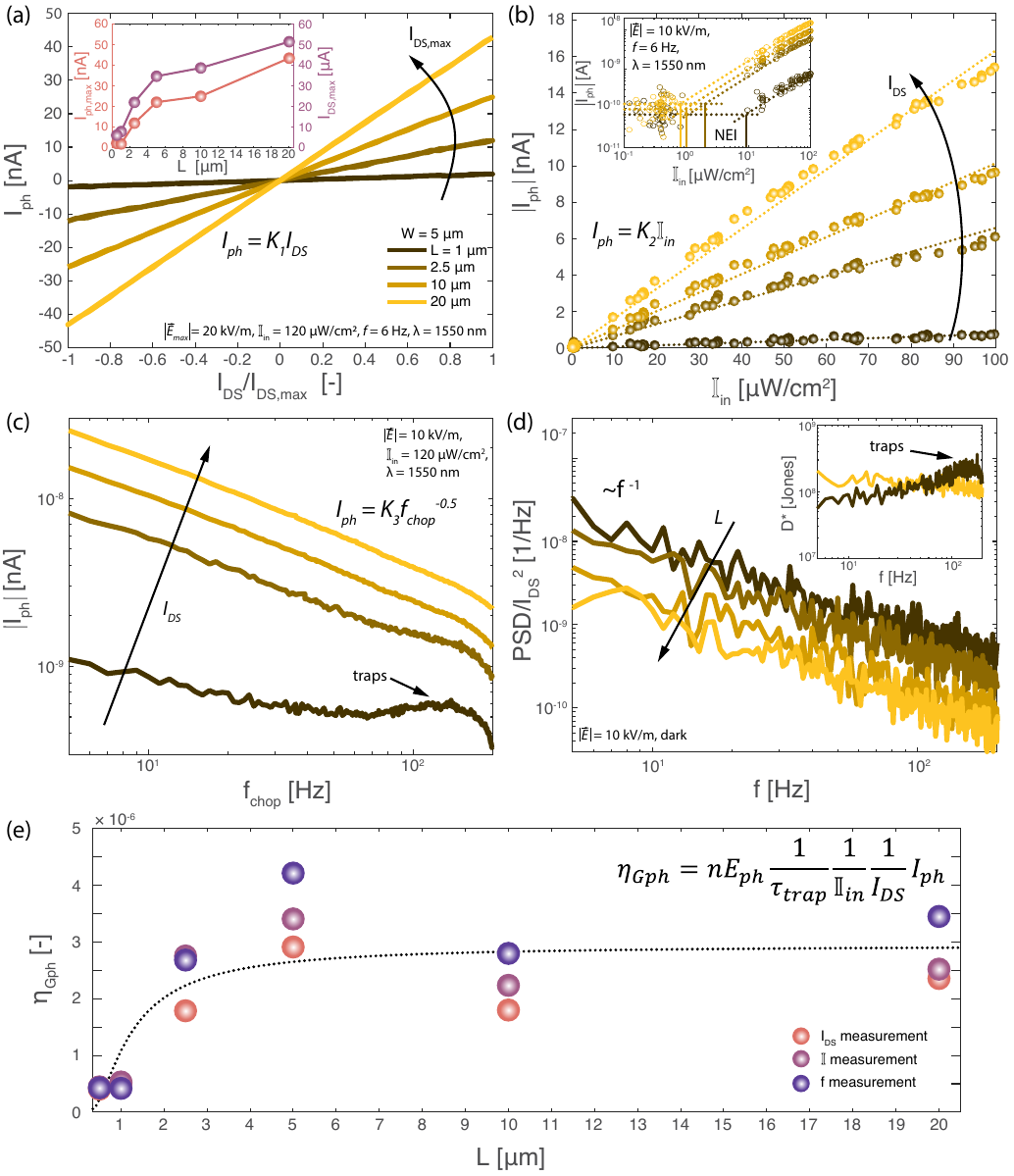}
	\caption{Bias, irradiance, and frequency dependence of photocurrent for a set of devices with $W = 5$ \unit{\um}, varying $L$, and \acs{LbL} spin coated \acp{cQD} PbS film of $t_{QDs} \approx 170$ \unit{nm}. (a) Linear depence of $I_{ph}=K_1 I_{DS}$. Bias voltage $V_{DS}$ range was scaled to reach a maximum of $\big|\vec{E}_{max} \big|=20$ \unit{kV \per m}. $I_{DS}$ was normalized by its maximum value for better illustration. The highest measured $I_{ph}$ and $I_{DS}$ for an extended set of devices are shown in the inset. (b) Linear dependence of $I_{ph} = K_2\mathds{I}_{in}$. The dotted lines show linear \acf{LSQ} through the data. Inset shows the \acf{NEI} in the log-log plot. (c) Frequency response of $I_{ph}$ by sweeping the light chopping frequency $f_{chop}$. $I_{ph}=K_3f_{chop}^{\alpha}$, with $\alpha=-0.5 \pm 0.1$ estimated from \acs{LSQ} fit to log-log plot. (d) \acf{PSD} of the noise current $I_{noise}$, measured in dark. \acs{PSD} was normalized by the measured $I_{DS}$. \acs{LSQ} fit the log-log plots show a mean exponent of -1.01 $\pm$ 0.02 to the frequency $f$. (e) Photogating efficiency $\eta_{Gph}$ estimate based on \acs{LSQ} fits giving proportionality factors $K_j$ from (a),(b), and (c). The full formula is given in the figure. $\tau_{trap}$ was estimated as $1/\sqrt{6}s$ in agreement with fits in (c). In (b),(c), and (d) $V_{DS}$ was scaled according to $\big|\vec{E} \big|=10$ \unit{kV \per m}. Note that measurements were performed on an extended set of devices including $L$ of 0.5, 1, 2.5, 5, 10, \qty{20}{\um}, but for clarity, only a distinct set is plotted in (a) to (d). Applied gate voltage of $V_G = 58$ \unit{V} for devices with $L$ of 0.5, 2.5, \qty{10}{\um} and $V_G=48$ \unit{V} for $L$ of 1, 5, \qty{20}{\um} for all the measurements.}
	\label{fig:ch_Scaling_Figure3}
\end{figure}

Next, the geometrical scaling of hybrid graphene – \acs{cQD} phototransistors was investigated. All devices were fabricated on the same sample, and the \acs{cQD} film was prepared by \acs{LbL} spin coating to reduce device-to-device variations. First, equation (\ref{eq:ch_Scaling_Iph}) was validated by bias, irradiance, and frequency sweeps for devices with different $L$ and a $W$ of \qty{5}{\um}. Figure \ref{fig:ch_Scaling_Figure3} (a) shows the linear relation between the source-drain current $I_{DS}$ and $I_{ph}$ with a proportionality factor $K_1$. The bias voltage was scaled according to $V_{DS} = \big|\vec{E} \big|L$ for comparable electric fields in the different devices. For better visualization, the measured source-drain current $I_{DS}$ was normalized by its highest value $I_{DS,max}$. The inset shows each device's highest photocurrents $I_{ph,max}$ at $I_{DS,max}$. Every device was measured at the gate voltage where the highest photocurrent was observed in the p-doped region of graphene (red dashed line in Figure \ref{fig:ch_Scaling_Figure2} (b)). The n-doped photocurrent maxima was experimentally not accessible as the \acs{LbL} spin coating led to Dirac point voltages up to \qty{70}{V}.

From equation (\ref{eq:ch_Scaling_Iph}), there is no length dependence expected, and all curves in Figure \ref{fig:ch_Scaling_Figure3} (a) should lay on a single line. The inset shows that despite the applied comparable electric field condition across devices, $I_{DS}$ still scales with $L$. This can be explained by a voltage drop over the contacts that becomes comparable to the voltage drop over the channel for smaller $L$, and thus $\big|\vec{E} \big|$ is not the same over the different channels despite the scaled $V_{DS}$. The contact resistance $R_C$ is estimated to be in the range of about \qty{0.6}{k\Omega} per contact by performed \acf{TLM} measurements. Thus, the ratio of $2\times R_C$ and the channel resistance $R_{Ch}$ become comparable ($2R_C/R_{Ch} \geq 0.5$) for devices with $L<5$ \unit{\um}. Reversibly, the contact resistance can be estimated from the spread of the $I_{ph}$ vs. $I_{DS}$ curves assuming the same $\big|\vec{E} \big|$ over the channel leads to the same $I_{DS}$. This also results in $R_C\approx0.6$ \unit{k\Omega}. With the fact that $I_{ph}$ is linearly proportional to $I_{DS}$, this confirms the origin of the $I_{ph}$ spread from contact limitations.

In Figure \ref{fig:ch_Scaling_Figure3} (b), irradiance sweeps were performed for the same set of devices. This $I_{ph}$–$\mathds{I}_{in}$ relation is linear with a proportionality factor $K_2$. $K_2$ is related to the responsivity by $R = \frac{K_2}{LW}$. As the curves follow a linear fit (dotted line), the detectors were operated in the linear dynamic regime ($R=const$), reaching responsivity values about \qty{300}{A\per W} for $\big|\vec{E} \big| = 10$ \unit{kV/m} and $f_{chop} = 6$ \unit{Hz}. The inset highlights the noise equivalent irradiance (NEI) for the different devices in the range of 1 to \qty{10}{\mu W\per cm^2}. The specific detectivity can be calculated by\cite{Huo2018,Koppens2014}
\begin{equation}
	\label{eq:ch_Scaling_Dstar}
	D^* = \frac{\sqrt{LW}\sqrt{\Delta f}}{NEP} = R \frac{\sqrt{LW}\sqrt{\Delta f}}{I_{noise}}.
\end{equation}
$NEP=NEI\times LW$ is the noise equivalent power, $\Delta f$ the frequency bandwidth, and $I_{noise}$ the noise current. The resulting detectivities are about $2\times 10^8$ \unit{Jones} at \qty{6}{Hz} and with $\Delta f=0.026$ \unit{Hz} (equivalent noise bandwidth of lock-in measurement).

Figure \ref{fig:ch_Scaling_Figure3} (c) shows the frequency response of the investigated devices. The light chopping frequency $f_{chop}$ was varied between 5 and \qty{200}{Hz}. A fit to data relates $I_{ph} = K_3 f^{\alpha}$ dependence, where $K_3$ is a proportionality constant. The curves follow $\alpha = -0.5$ ($\pm$ 0.1) up to frequencies of about \qty{160}{Hz}. A differing photocurrent dynamics can be observed for the \qty{1}{\um} channel length device (dark brown). Around \qty{70}{Hz}, the trend deviates from the $\alpha = -0.5$ as $I_{ph}$ increases with $f_{chop}$ before it drops eventually. 

The chopping frequency is defined over which period $\tau_{chop}$ the detector sees the light. Thus, it probes $\tau_{trap}$ of the \acs{cQD} film. Previous experiments have shown that at least two trapping times can be present underlying the photocurrent dynamics.\cite{Grotevent2021,Kara2023,Wang2015c} In analogy to low frequency $1/f$ noise\cite{Balandin2013}, the frequency response might originate from an envelope of individual Lorentzian shaped trap state contributions. Thus, the slope of the envelope $\alpha$ is a \acs{cQD} film-dependent quantity and depends on the specific film treatment as it varies across the literature (values between -0.1 to -2).\cite{Nikitskiy2016,Polat2019,Grotevent2021} Within this frame, a possible explanation of the rise of $I_{ph}$ around \qty{130}{Hz} might be the quenched escape rate from such a trap, as new carriers are excited before the trapped ones escape. Consequently, the traps might be constantly populated, and newly photoexcited e-h pairs can escape without being slowed down from those traps, leading to higher photocurrents. The energy depth of such a trap $i$ can be estimated by $\Delta E_{trap,i} = -kT ln(\tau_0/\tau_{trap,i})$, with $k$ the Boltzmann constant, and $T$ the temperature.\cite{Bube1967} The free charge carrier lifetime $\tau_0$ is the recombination time of a charge carrier contributing to the photocurrent, excluding the time it spends in a trap. For $1/\tau_{trap,i}=130$ \unit{Hz} an energy depth of $\Delta E_{trap,i} \approx 250$ \unit{meV} can be estimated, that corresponds to previously found surface trap state energies for PbS \acs{cQD} films between 100 to \qty{300}{meV}.\cite{Konstantatos2007a} For a spin coated film, this trap state is expected to be equally distributed throughout the film. However, in Figure \ref{fig:ch_Scaling_Figure3} (c), the trap appears for the device with a channel lengt of \qty{1}{\um}, and might be explained by the modulation of the trap in the vicinity of the source and drain electrodes.

Figure \ref{fig:ch_Scaling_Figure3} (d) shows the bias current normalized power spectral density of the measured noise current. The curves show a low frequency $1/f$-noise dependence as typically reported\cite{Balandin2013} for graphene FETs. The measurements were performed in the dark and at the same scaled bias condition $V_{DS} = \big|\vec{E} \big| L$. Fit to the curves reveal a mean exponent of -1.01 $\pm$ 0.02. The inset shows the derived specific detectivity $D^*$ using equation (\ref{eq:ch_Scaling_Dstar}) and taking the photocurrents from Figure \ref{fig:ch_Scaling_Figure3} (c) into account. The yellow line ($L=$ \qty{20}{\um}) has a near constant $D^*$ up to a frequency of \qty{200}{Hz} and represents most of the investigated devices. On the other hand, the dark brown curve ($L=$ \qty{1}{\um}) shows an increase of $D^*$ up to \qty{130}{Hz}, possibly due to quenched trapping sites leading to a rise of the photocurrent. The $D^*$ values of about $2 \times 10^8$ {Jones} fit well with the estimation from NEP found in Figure \ref{fig:ch_Scaling_Figure3} (b).

All the three introduced proportionality factors $K_j$ are related to $\eta_{Gph}$ and can be expressed from equation (\ref{eq:ch_Scaling_Iph}) as
\begin{equation}
	\eta_{Gph} = n E_{ph} \frac{1}{\tau_{trap}} \frac{1}{\mathds{I}_{in}} \frac{1}{I_{DS}} I_{ph}.
\end{equation}
The decay time $\tau_{trap}$ describes how long it takes to recover the dark current once the light is turned off. Thus, $\tau_{trap}$ defines the detector's maximum reachable photogain $G_{ph}$. To describe the experimentally observed $I_{ph}\sim f^{\alpha}$ dependence, and the fact that $f_{chop}$ probes $\tau_{trap,i}$, the chopping frequency can be modeled by $\tau_{trap} = \beta f_{chop}^{\alpha}$. Here, $\beta$ is introduced to formally correct the unit discrepancy arising from the fitting constant $\alpha$. As a result, the estimation of $\eta_{Gph}$ can be performed with $f_{chop}$, $\mathds{I}_{in}$, and $I_{DS}$ sweeps. Using the proportionality factors $K_j$ from  Figure \ref{fig:ch_Scaling_Figure3} (a) to (c), the photogating efficiency was estimated in Figure \ref{fig:ch_Scaling_Figure3} (d). $\eta_{Gph}$ yield a channel length dependence and similar values across the different measurements. Furthermore, the values of $\sim10^{-6}$ are compatible with the observation in Figure \ref{fig:ch_Scaling_Figure2} (c). A similar decreasing trend was observed for both $g_m$ and $\eta_{Gph}$ towards smaller $L$. Previously, contacting metals have been found to dope $\sim 500$ \unit{nm} into the graphene channel\cite{Xia2009}, thus reducing the effective gatable length of the channel, and limiting $g_m$ for small devices. As both $g_m$ (back gate) and $\eta_{Gph}$ (photogate) are related to the gating of the graphene channel, the same loss of gatable channel length might be the cause for decreasing $\eta_{Gph}$ with smaller $L$.

\begin{figure}[h!tb] 
	\centering
	\includegraphics[width=\linewidth]{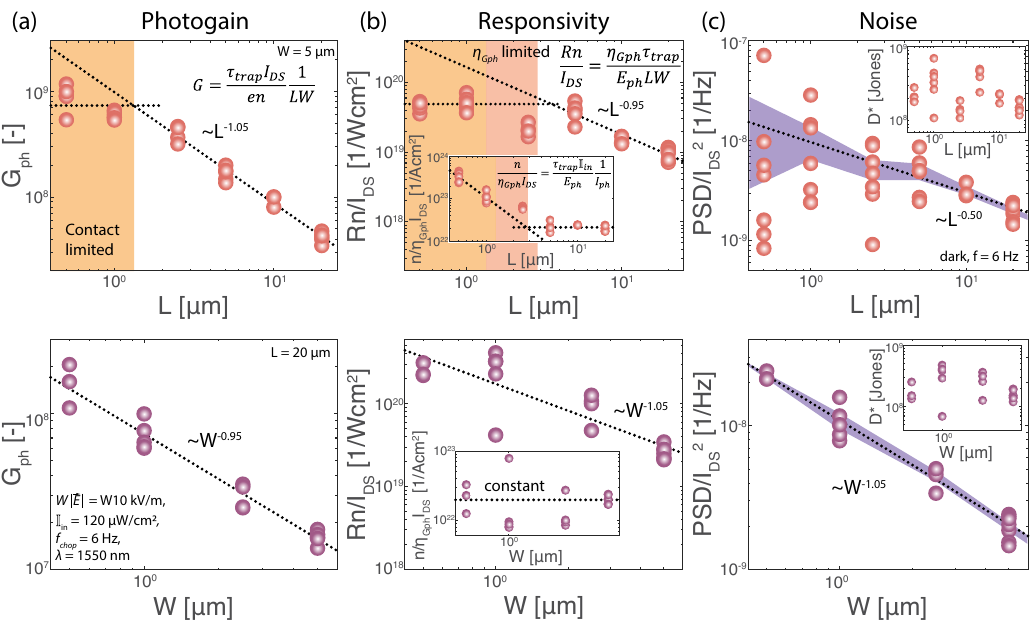}
	\caption{Length and width scaling of photoresponse. (a) photogain $G_{ph}$ scaling. \acs{LSQ} fits exponents for $L$ of 1.05$\pm$0.025 and for $W$ of 0.95$\pm$0.025 (dotted black line). The orange-shaded region highlights where the $L$-scaling breaks down as contact resistances become comparable to device channel resistances. (b) Responisivity $R$ scaling. $R$ was normalized by charge carrier density $n$ (different residual doping levels) and $I_{DS}$ (contact resistance). An exponent for $L$ of 0.95$\pm$0.05 and $W$ of 1.05$\pm$0.1 was fitted by \acs{LSQ}. The $L$-scaling shows contact limited (orange) and $\eta_{QE}$ limited range. The inset shows the $\eta_{QE}$ extended normalization factor for $R$, where the red-shaded region becomes visible. (c) \acf{PSD} of the noise current $I_{noise}$. The noise was measured in dark, and \acs{PSD} was normalized by the measured $I_{DS}^2$. \acs{LSQ} fits exponents of 0.502$\pm$0.02 for $L$-scaling and 1.05$\pm$0.03 (black dashed line). Purple shading shows the standard deviation of the mean values. Specific detectivity $D^*$ estimates between $10^8$ to $10^9$ are shown in the insets. The bias voltage $V_{DS}$ was scaled for a constant current condition according to $I_{DS} \sim W \frac{V_{DS}}{L}$. All values for (a) and (b) are extracted from maximum $I_{ph}$ points aquired in gate voltage sweeps, at a wavelength $\lambda=1550$ \unit{nm}, $\mathds{I}_{in}=120$ \unit{\mu W \per cm^2}, and $f_{chop} = 6$ \unit{Hz}. (c) was measured at the same $V_G$ values as in (a) and (b). The dotted horizontal lines indicate the mean values of the respective data sets.}
	\label{fig:ch_Scaling_Figure4}
\end{figure}

Next, the scaling of the photorespons by $L$ and $W$ was investigated as seen in Figure \ref{fig:ch_Scaling_Figure4}. For that devices with $L$(number of devices) of 0.5 ($\times$5), 1 ($\times$6), 2.5 ($\times$4), 5 ($\times$5), 10 ($\times$3), and \qty{20}{\um} ($\times$7) with a width of $W=$ \qty{5}{\um} were measured for the $L$-scaling. For $W$-scaling $W$(number of devices) of 0.5 ($\times$3), 1 ($\times$5), 2.5 ($\times$4) and \qty{5}{\um} ($\times$7) with a $L=$ \qty{20}{\um} were characterized. All the devices were on the same chip and had identical \acs{cQD} film properties, allowing for a cross-comparison. The quantities $I_{ph}$, $I_{DS}$, $n$ where extracted from gate sweeps as described in Figure \ref{fig:ch_Scaling_Figure2} (a). The values were taken at respective maximum photocurrent points in the p-doped region of graphene. As the photoresponse is proportional to $I_{DS}$, $V_{DS}$ was scaled for a comparable current condition according to $I_{DS} \sim W \frac{V_{DS}}{L}$.

In Figure \ref{fig:ch_Scaling_Figure4} (a), $G_{ph}$ scaling by $L$ (top) and $W$ (bottom) is presented. A fit to the data (black dotted line) shows the $1/L$ trend and agrees with the prediction from equation (\ref{eq:ch_Scaling_GainIds}). The horizontal dotted line indicates the mean value in that range. The scaling law breaks down around \qty{1.5}{\um} (orange shaded region), where the total graphene FET resistance $R_T$ is about \qty{2}{k\Omega}. This is comparable to the contact resistance of \qty{1.2}{k\Omega} (\qty{600}{\Omega} per contact) and thus points to a current injection limitation at the specified biasing condition. The $W$-scaling follows a predicted $1/W$ trend over the evaluated region. As the $W$-scaling devices have a $L=$ \qty{20}{\um}, the smallest resistance for the $W = 5$ \unit{\um} is about \qty{8}{k\Omega} such that no current injection limitation is observed.

Figure \ref{fig:ch_Scaling_Figure4} (b) depicts the scaling of photoresponsivity $R$. The responsivity was normalized by $n/I_{DS}$ to account for device-to-device variations (different residual doping levels) and contact resistance. $R$ follows also a $1/L$-scaling law. Although $I_{ph}$ is independent on geometry for constant $\mathds{I}_{in}$, $P_{in} = \mathds{I}_{in}LW$ and thus 
\begin{equation}
	R = \frac{I_{ph}}{\mathds{I}_{in}LW}.
\end{equation}
The breakdown of the scaling law, however, exceeds the current injection limited region caused by contacts (yellow shaded) and extends up to $\sim$ \qty{3}{\um}. The inset shows the normalization factor additionally accounting for $\eta_{Gph}$. As $\eta_{Gph}$ is introduced, the red shaded area appears, thus being responsible for an additional length limitation to the scaling. The bottom panel shows the same normalized $R$ for the $W$-scaling. In fact, as for the gain, there is a good agreement for the $1/W$ scaling observed over the investigated width scale. Moreover, the normalization factor, including $\eta_{Gph}$, shows a constant trend. This highlights that $\eta_{Gph}$ might be similarly affected by reduced gatable channel lengths due to metal doping, as mentioned for $g_m$ above.

Finally, the influence of scaling on the noise and specific detectivity $D^*$ was investigated in Figure \ref{fig:ch_Scaling_Figure4} (c). From $I_{DS}^2$ normalized \acs{PSD} spectra as shown in Figure \ref{fig:ch_Scaling_Figure3} (d), the values at \qty{6}{Hz} were extracted for every device. The measurements were performed in dark and at the same gate voltage where the maximum photocurrent was extracted previously. The dotted lines show fits through the mean values of each length or width, whereas the purple shading indicates the standard deviation. The fits suggest a $1/\sqrt{L}$ and $1/W$ dependence for the characterized devices. For the $L$-scaling, the distribution of points increases as $L$ gets smaller, whereas the $W$-scaling has an overall smaller spread of the data. The inset shows derived $D^*$ about \qty{3e8}{Jones}, using the photoresponse shown in (b).

Low frequency $1/f$ noise in \acsp{GFET} is primarily attributed to electrostatic fluctuations arising from trap states at the oxide interface and leads to conductivity fluctuations in the channel.\cite{Balandin2013,Karnatak2017} Conductivity $\sigma$ is related to the product of $n\mu$. Hence, the two main models that describe $1/f$ noise are based on $\Delta n$ (McWorther's model) or $\Delta \mu$ (Hooge's law) fluctuations.\cite{Simoen1999,Balandin2013} Both models predict a $1/LW$ dependence of noise in the \acs{FET} channels.\cite{Simoen1999} Accordingly, a volume scaling was experimentally verified for metallic break junctions to hold in the diffusive transport regime.\cite{Wu2008d} For graphene FETs, however, the $1/LW$ scaling has been found to experimentally not describe the data in full detail.\cite{Rumyantsev2010} Being restricted to a surface only, graphene is very sensitive to its environment. Thus, perturbations to the channel such as bilayers\cite{Heller2010}, contacts\cite{Karnatak2016,Rumyantsev2010}, and environmental exposure\cite{Rumyantsev2010} are altering the electrostatic screening potential of the channel and influence the low-frequency $1/f$ noise. Like the photoresponse, though, reducing $L$ is pronouncing the contact effects. Hence, an increasing spread for smaller $L$ channels is observed. In addition, possible regions of double layers due to non-ideal CVD graphene growth and the additional current path through the \acs{cQD} can thus be the reason for a deviation from an overall $1/LW$ trend observed in the \acp{PSD}. 

Expanding the demonstrated photocurrent characteristics of equation (\ref{eq:ch_Scaling_Iph}) into the detectivity yields
\begin{equation}
	\label{eq:ch_Scaling_DstarIph}
	D^*=\frac{1}{n}\frac{1}{E_{ph}}\tau_{trap}\eta_{QE}\frac{1}{\sqrt{LW}}\frac{1}{\sqrt{PSD/I_{DS}^2}}.
\end{equation}
If now $\sqrt{PSD/I_{DS}^2}\sim1/\sqrt{LW}$, there is no geometrical dependence expected in $D^*$, which is consistent with the performed measurements. It is important to mention that the experimentally determined $D^*$ values between $10^8$ to $10^9$ \unit{Jones} are around four orders of magnitude lower than the highest values found in the literature\cite{Konstantatos2012} of $10^{13}$ \unit{Jones}. The difference can be explained mainly by two factors from equation (\ref{eq:ch_Scaling_DstarIph}). First, the difference is in the found $\eta_{Gph}$ range. As $\eta_{Gph}$ does not differ much for devices with \acs{EHD} printed (single step ligand exchange) and a \acs{LbL} spin coated \acs{cQD} films, the interface between graphene and \acs{cQD} film limits most likely the charge transfer. CVD graphene processing is prone to defects and residuals from the growth and device fabrication. On the contrary, the highest $D^*$ values were reached with exfoliated single-crystal graphene. Second, $\tau_{trap}$ defines the maximum reachable gain and is related to the photocurrent dynamics ($I_{ph}$ to $f_{chop}$ relation). Preserving the $I_{ph}$ magnitude for increasing $f_{chop}$ leads to increasing $D^*$ due to lower $1/f$ noise in the channel as demonstrated in Figure \ref{fig:ch_Scaling_Figure3} (c) and (d).

\subsection*{Conclusion}

Here, the scaling of hybrid graphene-PbS \acs{cQD} phototransistors with channel length $L$, width $W$, and \acs{cQD} film thickness $t_{QDs}$ was experimentally demonstrated. The photocurrent $I_{ph}$ was found to be independent of $L$ and $W$ but increased with $t_{QDs}$. Instead, the photogain $G_{ph}$ and responsivity $R$ scaled with $1/LW$. Although a $1/L^2$ dependence can be derived for $G_{ph}$, this neglects increasing channel currents with $I_{DS}\sim V_{DS}W/L$. Relating $G_{ph}$ to $I_{DS}$ leads to the $1/LW$ relation. For $R$, on the contrary, the geometrical dependence is evolving from a constant irradiance $\mathds{I}_{in} = P_{in}/LW$, rather than a constant light power $P_{in}$ seen by the detector. The $L$ scaling broke down for $L<3$ \unit{\um} due to contact metal contributions (contact resistance and channel doping). Although specific detectivity is $D^*\sim R$, the found values in the range between $10^8$ to $10^9$ \unit{Jones} ($\lambda=1550$ \unit{nm},  $f = 6$ \unit{Hz}, room temperature) were independent on geometry. This is due to increasing noise currents for smaller dimensions. The photoresponse dependency on back gate voltage $V_G$, bias $I_{DS}$, irradiance $\mathds{I}_{in}$, and light chopping frequency $f_{chop}$ were analyzed. A derived photogating efficiency $\eta_{Gph}$ of $\sim10^{-6}$ from those measurements pointed to a \acs{cQD}-graphene interface limited charge transfer as a source for further photoresponse improvement.

\subsection*{Methods}

\begin{footnotesize}
	
	\paragraph{Graphene Growth} ~\\
	CVD graphene was grown on commercial copper foil (Foil 2017, No. 46365, Alfa Aesar). The Cu-foil was first sonicated in acetone (\qty{15}{min}) and rinsed with deionized (DI) water. The foil was then immersed in nitric acid HNO\textsubscript{3} (\qty{30}{min}), twice sonicated in DI-water (\qty{1}{min}), immersed in ethanol (\qty{1}{min}), and blow-dried with N\textsubscript{2}. The Cu-foil was annealed in H$_2$ (\qty{20}{sccm} H$_2$ and \qty{200}{sccm} Ar mixture) at \qty{1000}{\degreeCelsius} and a pres-sure of $<$ \qty{1}{mbar} for \qty{70}{min} in a furnace. The graphene was grown from \qty{0.04}{sccm} methane CH\textsubscript{4} for \qty{21}{min} at \qty{1000}{\degreeCelsius} and a pressure of \qty{110}{mbar}, subsequentyl cooled down by opening the temperature shielding.
	
	\paragraph{\Acl{GFET} Fabrication} ~\\
	A p-Si substrate with \qty{285}{nm} chlorinated dry thermal SiO\textsubscript{2} was prepatterned with contacts. The contacts were fabricated by e-beam-assisted thermal evaporation of Ti/Au (5/40 \unit{nm}) on a photolithographically defined resist (AZ5214E). PMMA 50K-protected graphene was prepared for wet transfer by Cu-etching in commercial Transene etchant (1 h), rinsed with DI water, left floating on hydrochloric acid (5 min), and re-rinsed with DI water. Graphene was transferred on the pre-patterned p-Si substrate from DI water, dried on a hotplate, and put into a vacuum ($\sim$\qty{60}{h}). Graphene channels were O\textsubscript{2}-plasma etched using a Cu-mask, followed by Cr/Au top-contacts (2/40 \unit{nm}), both defined by e-beam lithography (PMMA 50K/PMMA 950K).
	
	\paragraph{PbS \acs{cQD} Film Fabrication} ~\\
	PbS QDs were synthesized following Hines et al. \cite{Hines2003} with slight adaptations as described in detail previously\cite{Kara2023}. For \acf{EHD}, the PbS \acs{cQD} were redispersed in n-tetradecane with a concentration of \qty{40}{mg \per ml} and filtered with \qty{0.1}{\um} PTFE syringe filter. Printing nozzles with openings between 2 to 3 \unit{um} were fabricated by pulling glass capillaries (World Precision Instruments TW100-4) and subsequent coating Ti/Au (5/\qty{50}{nm}). For the printing, the nozzle was brought in $\sim 5$ \unit{\um} proximity to the sample and an AC voltage of \qty{280}{V} (\qty{250}{Hz}) was applied between sample and nozzle. The structures were printed with a stage speed of \qty{6}{\um \per s}, line-spacing about \qty{500}{nm}, and subsequent overprinting was oriented in a staggered fashion. The EHD printing system used here was previously described in more detail.\cite{Antolinez2019} A single-step ligand exchange was performed by overnight soaking the \acs{cQD} films in 2 vol\% \acf{EDT} in acetonitrile. Subsequently, the samples were rinsed in acetonitrile and dried with N\textsubscript{2}.
	
	For the \acf{LbL} prepared film, the PbS \acsp{cQD} were redispersed in octane with a concentration of \qty{20}{mg \per ml}. A layer of PbS \acsp{cQD} was spin-coated (\qty{2500}{rpm}, \qty{45}{s}), subsequently a drop of 2 vol\% \acf{EDT} in acetonitrile was placed for \qty{30}{s} before spinning the sample dry, and followed by one drop of acetonitrile and one drop of octane while sample was spinning (\qty{2500}{rpm}, \qty{45}{s}). The steps were repeated six times for a $\sim 170$ \unit{nm} film thickness.
	
	Before film fabrication, the PbS \acsp{cQD} were characterized by UV-Vis spectroscopy (Jasco V-670). The fabricated PbS \acs{cQD} films were analyzed by AFM (Bruker Dimension ICON 3), and the device cross-section was investigated by FIB-SEM (FEI Helios NanoLab G3 UC).
	
	\paragraph{Device Characterization} ~\\
	The \acs{GFET} were electrical characterization before PbS \acs{cQD} film deposition in a two-probe configuration (Keithley 4200-SCS semiconductor characterization system). The field-effect mobility $\mu = \frac{L}{W V_{DS} C_G} \frac{dI_{DS}}{dV_{G}}$ was estimated by linear \acf{LSQ} to transfer curves ($I_{DS}$ vs. $V_G$). The contact resistance was estimated by the transfer length method (TLM).
	
	To characterize the photoresponse, a broadband light source (Thorlabs, SLS201) was modulated with a chopper (Thorlabs, MC2000B-EC) and focused onto a monochromator (Princeton Instruments, SpectraPro HRS-300 spectrometer with grating 150 G/mm, blaze 0.8 \unit{\um}). Subsequent long-pass filters attenuated the higher spectral orders (400, 600, 800, 1200, 1900 nm) at the monochromator exit. The light was collimated with a lens and split with a 50/50 Polkadot beamsplitter. Onto one end, a reference detector (Gentec, UM-9B-L) was placed, and the light irradiance was determined by a lock-in amplifier (Stanford Research System, SR865A). Onto the other light path end, the samples were placed into an optically accessible cryostat (JANIS ST-100) with a quartz glass window. Gate voltage and source-drain bias were applied with SMUs (Keithley, 2614B and 2450). The photovoltage was measured over a shunt resistance of \qty{1}{k\Omega} with lock-in amplifiers (Stanford Research System, SR860).
	
	Noise currents were measured in the dark with a battery-powered trans-impedance amplifier (Stanford Research Systems, SR570). The signal was low pass filtered (10 kHz) and subsequently acquired with a data acquisition board (National Instruments, USB6341) at a sampling rate of 500 kHz. The DC offset was removed for the power spectral densities estimates, and $10\times$ one-second-long time traces were averaged.
	
	All measurements were performed at room temperature and vacuum (4$\cdot 10^{-7}$ mbar). The line frequency of \qty{50}{Hz} and its harmonic \qty{150}{Hz} were removed from all the frequency-dependent measurements. The measurements were repeated throughout the study duration for a selected set of devices ($\sim21$ \unit{days}). In between the measurements, the samples were taken out of the measurement setup, rebonded in atmosphere, and placed back in the measurement setup. The measurements were reproducible with a maximal deviation of $\sim$ 30\%. A device of each dimension was measured before the next one with the same dimension was investigated to reduce the influence of possible variation from sample alignment or degradation on the study outcome.
\end{footnotesize}

\subsection*{Acknowledgements}
\begin{footnotesize}
	The authors thank FIRST-Lab (Center for Micro- and Nanoscience) at ETH Zurich for access to the clean-room and the Swiss National Science Foundation (SNSF, project no. 200021 182790) for financial support.
\end{footnotesize}

\subsection*{Author Contribution}
\begin{footnotesize}
	G.K. and I.S. conceived the study and planned the experiments. G.K. fabricated the samples, performed the measurements and analyzed the data. P.R. performed the EHD printing with supervision by D.P. E.W. performed the FIB-SEM images. D.D. synthesized PbS colloidal QDs. R.F. grew graphene. G.K., I.S. and M.C. discussed the data. I.S., M.C., M.K, D.P. initated and supervised the project. G.K. wrote the manuscript with inputs and discussions from all authors.
\end{footnotesize}

\FloatBarrier
\clearpage



\newpage
\scriptsize   
\bibliographystyle{ieeetr}
\bibliography{./Bibliography/library}

\end{document}